%Paper: hep-ph/9211227
%From: FORTE@TO.INFN.IT
%Date: Sat, 7 NOV 92 23:53 GMT
%Date (revised): Tue, 17 Nov 1992 17:40:52 +0100 (WET)

\magnification=1200
\hoffset=1truecm
\def\uplrarrow#1{\raise1.5ex\hbox{$\leftrightarrow$}\mkern-16.5mu #1}
\def\bx#1#2{\vcenter{\hrule \hbox{\vrule height #2in \kern #1\vrule}\hrule}}

\def\squiggle#1{\lower1.5ex\hbox{$\sim$}\mkern-14mu #1}%I changed 16 to 14?

\def\narrower{\advance\leftskip by\parindent \advance\rightskip by\parindent}

\def\mbox#1#2{\vcenter{\hrule width#1in\hbox{\vrule height#2in
   \hskip#1in\vrule height#2in}\hrule width#1in}}
\def\eqsquare #1:#2:{\vcenter{\hrule width#1\hbox{\vrule height#2
   \hskip#1\vrule height#2}\hrule width#1}}
\def\inbox#1#2#3{\vcenter to #2in{\vfil\hbox to #1in{$$\hfil#3\hfil$$}\vfil}}
       %margin bullet
\def\strutdepth{\dp\strutbox}
\def\marbul{\strut\vadjust{\kern-\strutdepth\specialbul}}
\def\specialbul{\vtop to \strutdepth{
    \baselineskip\strutdepth\vss\llap{$\bullet$\qquad}\null}}
\def\Bcomma{\lower6pt\hbox{$,$}}    % Big commutator
\def\bcomma{\lower3pt\hbox{$,$}}    % commutator

\def\updots{\mathinner{\mskip 1mu\raise 1pt\hbox{.}
    \mskip 2mu\raise 4pt\hbox{.}\mskip 2mu
    \raise 7pt\vbox{\kern 7pt\hbox{.}}\mskip 1mu}}

\def\pmb#1{\setbox0=\hbox{#1}%
     \kern-.025em\copy0\kern-\wd0
     \kern.05em\copy0\kern-\wd0
     \kern-.025em\raise.0433em\box0}

\def\1{\;1\!\!\!\! 1\;}

\def\m@th{\mathsurround=0pt}
\def\upsquarefill{$\m@th\bracelu\leaders\vrule\hfill\braceru$}
\def\ope#1{\mathop{\vtop{\ialign{##\crcr
     $\hfil\displaystyle{#1}\hfil$\crcr\noalign{\kern3pt\nointerlineskip}
     \kern4pt\upsquarefill\kern4pt\crcr\noalign{\kern3pt}}}}\limits}

\def\lsim{\mathrel{\rlap{\lower4pt\hbox{\hskip1pt$\sim$}}
    \raise1pt\hbox{$<$}}}         %less than or approx. symbol
\def\gsim{\mathrel{\rlap{\lower4pt\hbox{\hskip1pt$\sim$}}
    \raise1pt\hbox{$>$}}}         %greater than or approx. symbol

\vsize=7.5in
\hsize=5in
\pageno=0
\tolerance=10000
\hyphenation{Pre-daz-zi}

\hfuzz=5pt
\baselineskip 12pt plus 2pt minus 2pt
\hfill DFTT 64/92

\hfill  October 1992
\medskip
\centerline{\bf VIOLATION OF THE BJORKEN SUM RULE IN QCD}
\vskip 36pt\centerline{Stefano
Forte}
\vskip 12pt
\centerline{\it I.N.F.N., Sezione di Torino}
\centerline{\it via P.~Giuria 1, I-10125 Torino, Italy}
\vskip 1.5in
{\narrower\baselineskip 12pt
\centerline{\bf ABSTRACT}
\noindent We show that the Bjorken sum rule for the first moment of
the polarized nucleon structure function $g_1$ is not necessarily
satisfied in QCD if
the axial anomaly is taken into account. We argue
that nonperturbative  QCD effects should lead to a violation of the sum
rule of the order of the percentage isospin violation in current quark
masses.}

\vskip 1.2in
\centerline{Submitted to: {\it Physics Letters B}}
\vfill
\eject
\input harvmac
%\draftmode
\baselineskip 24pt plus 4pt minus 4pt
The Bjorken sum rule \ref\bj {J.~D.~Bjorken,
{\it Phys. Rev} {\bf 148}, 1467 (1966)},
which relates the isotriplet component of the
first moment of the polarized nucleon structure function $g_1$ to the
weak couplings $G_A$ and $G_V$ is usually considered a solid consequence
of the standard model, and has been traditionally viewed as
a benchmark
of current universality and the quark interpretation. Even though this
sum rule has never been tested directly, it is routinely used in the
analysis and extraction of polarized structure function data.

Recent
high-precision experimental determinations of first moments of
polarized \ref\emc{J.~Ashman {\it et al.}, {\it Nucl. Phys.} {\bf B328}, 1
(1990)} (for a discussion
see \ref\alta{G.~Altarelli, in `The Challenging Questions'',
Proceedings of the 1989 Erice school, A.~Zichichi, ed. (Plenum, NY, 1990)})
and unpolarized \ref\nmc{P.~Amaudruz {\it et al.}, {\it Phys. Rev. Lett.}
{\bf 66}, 560 (1991)} (for a discussion see \ref\megott{S.~Forte,
Torino preprint DFTT 18/92 (1992), {\it Phys. Rev. D},
in press}) structure functions have lead to results which disagree
with naive expectations based on partonic intuition, and have lead to
reconsider assumptions, such as the isoscalarity of the antiquark sea in
the nucleon, which, even though not a consequence of QCD, were always
made in the analysis of experimental data.
It is thus reasonable to ask,
what is the status of the Bjorken sum rule in QCD? Two recent statements
which express the  common
wisdom of, respectively, the experimental and theoretical
communities in this respect are ``today the Bjorken sum rule is regarded as
a test of QCD''\ref\pres{C.~Y.~Prescott, SLAC Preprint SLAC-PUB-5604 (1991)},
and ``the Bjorken sum rule presents a basic test of the current
algebra''\ref\drell{S.~D.~Drell,  SLAC Preprint SLAC-PUB-5720 (1992)}.

Here
we shall show that, whereas the latter statement is correct, the former is not:
it is possible to violate the Bjorken sum rule without violating
QCD --- nor its current algebra, which, due to the U(1) problem, is not the
same as the quark current algebra \ref\uone{See e.g. S.~Coleman, ``Aspects
of Symmetry'' (Cambridge, 1985)}.
In particular, due to a
gluonic admixture
to matrix elements of the
singlet axial current, triggered by the axial anomaly,
the Bjorken sum rule is only true provided one makes an assumption of
isoscalarity of the gluon sea. This assumption, although plausible, need
not be true, and actually, we shall argue
that non-perturbative gluon
configurations (like instantons) should violate the assumption if the (light)
current quark masses are unequal (as they are in the real world).

Let us first review the standard derivation of the Bjorken sum rule.
Both current-algebra arguments \drell\ and the operator-product
expansion \alta\ lead to equate the first moment of the polarized structure
function $g_1$ with the (forward)
matrix element of the electromagnetic axial current:
\eqn\ope
{\eqalign{\int \! dx \, g_1(x) &=C A_1\cr
A_1s^\mu&\equiv\langle N| \sum_i q^2_i \bar\psi_i\gamma^\mu\gamma_5\psi_i
|N \rangle ,
\cr}}
where $s^\mu$ is the spin four-vector of the given hadronic target $N$,
$s_\mu s^\mu=-1$, and the sum runs over quark flavors with
charge $q_i$.
The coefficient $C$ is computed perturbatively as
$C={1\over2}\left(1-{\alpha_s\over \pi}+ O(\alpha_s^2)\right)$.
In order to extract a useful sum rule from Eq.\ope\ one takes an isotriplet
matrix element, i.e., the proton-neutron difference
\eqn\trip
{\int \! dx \, \left(g_1^p(x)-g_1^n(x)\right)
=-s_\mu C \langle N| \sum_i q^2_i \bar\psi_i\gamma^\mu\gamma_5\psi_i
|N \rangle^{I=1} ,
}
then one uses current universality  to express the isotriplet matrix element on
the r.h.s. of Eq.\trip\ as the matrix element of the isotriplet part of the
current, i.e.,
\eqnn\srule\eqnn\isoe
$$\eqalignno{\int \! dx \, \left(g_1^p(x)-g_1^n(x)\right)
&=-s_\mu C \langle N|
\left(\sum_i q^2_i \bar\psi_i\gamma^\mu\gamma_5\psi_i\right)^{I=1}
|N \rangle&\srule\cr
&=C{1\over 3}{G_A\over G_V},&\isoe\cr
}$$
where the last equality --- the Bjorken sum rule ---
which relates the matrix element of the isotriplet
current to the weak couplings $G_A$, $G_V$,
follows form straightforward isospin algebra.

Whereas the derivation of Eq.\ope\ is based on QCD current algebra and
asymptotic freedom, and thus a failure of that equation would surely undermine
QCD, in order to get to the Bjorken sum rule proper \isoe\ one needs
the extra step leading from Eq.\trip\ to  Eq.\srule. Even though this seems
quite innocuous, the assumption on which is based taken at face value is
obviously false. Indeed, whereas the quark currents form a
U$_{\rm V}$(3)$\times$U$_{\rm A}$(3) current algebra, at the hadronic
level this is broken down to
U$_{\rm V}$(3)$\times$SU$_{\rm A}$(3)
because of the axial anomaly which removes the singlet axial
U$_{\rm A}$(1)
symmetry \uone.
This is manifested, for instance, by the fact that the matrix elements of the
singlet current are scale dependent, while those of the nonsinglet one aren't.
It follows that the singlet current's matrix elements are not in the same
multiplet as those of the nonsinglet ones, thus we do not know their
transformation properties under isospin.

We can see this more explicitly by introducing a decomposition \alta,
\ref\mespin\ {S.~Forte, {\it Phys. Lett.} {\bf B224}, 189 (1989);
{\it Nucl. Phys. } {\bf B331}, (1990)}
of the form factor $G_A$ of the forward nucleon matrix
element of the isosinglet axial current
in terms of its
quark components $\Delta q_i$  and gluon component $\Delta G$:
\eqn\twocomp
{\eqalign{G_A s^\mu\equiv&\langle N|\sum_i \bar\psi_i\gamma^\mu
\gamma_5\psi_i|N\rangle\cr
G_A&=\sum_i\left( \Delta q_i - \Delta G\right)\cr
\Delta& G=\left(\Omega+ \Delta g\right),\cr}}
where $\Delta g$ and $\Omega$ are, respectively, the contributions from
perturbative and non-perturbative gluonic configurations to $\Delta G$.
Despite some controversies on the way the gluonic
contribution could be experimentally singled out, it is now
clear \ref\althera\
{See G.~Altarelli, CERN preprint CERN-TH.6340/91 (December 1991)} that the
decomposition \twocomp\ can be established in a gauge-invariant way both
perturbatively in the
QCD parton model \alta\ and nonperturbatively at the operator
level \mespin. The simplest way of introducing it is
to observe \mespin\ that the matrix element of the singlet axial charge
$Q_5\equiv\int d^3x \sum_i  \bar\psi_i\gamma^0\gamma_5\psi_i$ is, in general,
the sum of the canonical quark helicity operator $Q_5^q$ and a gluonic operator
$Q_5^g$, whose nucleon matrix elements are respectively identified with $\Delta
q$ and $\Delta G$. Whereas $Q_5^q$ is conserved, $Q_5^g$ is not and the
gluonic contribution is thus responsible for the anomalous divergence of the
axial current\ref\anom{See R.~Jackiw,
in S.~B.~Treiman, R.~Jackiw, B.~Zumino and E.~Witten,
``Current Algebra and Anomalies'' (World Scientific, Singapore, 1985)}.
The separation of the gluonic contribution in
a perturbative and a nonperturbative portion
is discussed in detail in Ref.\mespin,
and can be understood \althera\ observing
that the (perturbative) gluon distribution
provides a contribution $\Delta g$ to $\Delta G$ which evolves at two loops,
while  nonperturbative field configurations (such as instantons) provide a
scale-invariant contribution.

Using Eq.\twocomp\ in Eq.\trip, neglecting QCD corrections,
and assuming strange and heavier flavor
distributions to be isosinglet to good approximation (we shall come back on
this later) we get
\eqn\corr
{\eqalign{&\int \! dx \, \left(g_1^p(x)-g_1^n(x)\right)
=\cr
&\qquad={1\over 2} \left[-s_\mu\langle N| \left({4\over9}
\bar u\gamma^\mu
\gamma_5u+{1\over9}\bar d\gamma^\mu
\gamma_5 d\right)
|N \rangle^{I=1} -{5\over9}\langle N| \Delta G |N \rangle^{I=1} \right].
\cr}}
The (non-anomalous) quark
contributions $\Delta q$ to $G_A$ do belong in the full
U(3)$\times$U(3)
multiplet, thus we can use isospin symmetry to relate $\Delta u^p=\Delta d^n$
and so forth (where the superscript denotes the state in which the matrix
element is taken) to get
\eqn\newbj
{\eqalign{\int \! dx \, \left(g_1^p(x)-g_1^n(x)\right)
&={1\over 2}\left[{1\over3} \left(\Delta u^p - \Delta d^p\right)
- {5\over9}\left(\Delta G^p-\Delta G^n\right)\right]\cr
&={1\over 2}\langle N|\left(\sum_i q^2_i \bar\psi_i
\gamma^\mu\gamma_5\psi_i\right)^{I=1}|N\rangle-
{5\over18}\left(\Delta G^p-\Delta G^n\right)\cr
&=C\left[{1\over 3}\left({G_A\over G_V}\right)
-{5\over9}\left(\Delta G^p-\Delta G^n\right)\right].\cr
}}
It follows that the Bjorken sum rule is corrected at the crucial step \srule\
by the last term on the r.h.s. of Eq.\newbj, and it is true only upon the
assumption that this term vanishes.

We have thus shown explicitly that the Bjorken sum rule follows from QCD only
upon the assumption that the gluonic component $\Delta G$ is isosinglet.
Because there are now reasons to believe that $\Delta G$ may be actually rather
large (i.e., comparable to $\Delta q_i$) \alta, it is important to see
whether we should expect indeed $\Delta G$ to be isosinglet or not. It should
be realized that since $\Delta G$ receives in general both perturbative
and nonperturbative contributions, there is no {\it a priori} reason why it
should be isosinglet. Rather, this is a phenomenological assumption on the
same footing as that according to which the nucleon sea should be SU(2)
symmetric, which has recently been disproved experimentally \nmc,\megott.

Indeed, isospin violations of order 5\% in the sea parton distributions
are suggested \megott\ by recent
data \nmc, and isospin violation of, for example, the nucleon radii up to
about 0.2 Fm are compatible with current electron-nucleus scattering data
\ref\iso{T.~W.~Donnelly, J.~Dubach and I.~Sick, {\it Nucl. Phys.} {\bf A503},
589 (1989)}. Due to the numerical coefficients in Eq.\newbj\
assuming \alta,\mespin $|\Delta G|\sim{1\over N_f} \Delta q$,
this leads to expect  violation of the Bjorken sum rule of order 2 -- 3\%
just from isospin violation in the gluon distributions. If $\Delta G$ receives
a sizable nonperturbative contribution, however, even larger violations of the
Bjorken sum rule may appear.

To understand this, it is convenient to consider the matrix
element of the divergence of the (singlet) axial current, which is given by
the anomaly equation \anom:
\eqn\anoma
{G_A s\cdot (p^\prime -
p)\equiv \langle N |\partial_\mu j^\mu_5 | N \rangle= \langle N |
\sum_i 2 m_i\bar\psi i\gamma_5\psi + {N_f\over 8\pi^2}g^2
\epsilon^{\mu\nu\rho\sigma} F_{\mu\nu} F_{\rho\sigma}| N \rangle,}
where $(p^\prime - p)$ is the momentum transfer and
$m_i$ is the current quark mass for the $i$-th flavor.
In order to avoid possible sources of confusion, it should
be pointed out rightaway that
the naive identification of the two terms on the r.h.s. of
Eq.\anoma\ with the quark and gluon contributions $\Delta q$ and $\Delta G$
is incorrect \ref\vene{G.~M.~Shore and G.~Veneziano {\it Phys. Lett.}
{\bf B244}, 75 (1990); {\it Nucl. Phys.} {\bf B381}, 3 (1992)}.
Next, it is convenient to define \ref\gtw{D.~G.~Gross, S.~Treiman and F.
{}~Wilczeck, {\it Phys. Rev.} {\bf D19}, 2188 (1977)}\ref\venold{G.~Veneziano,
{\it Mod. Phys. Lett.} {\bf A4}, 1605 (1989)}, \vene (see also
Ref.\ref\liu{K.~F.~Liu, {\it Phys. Lett.} {\bf B281}, 141 (1992)})
one-meson reducible and
one-meson irreducible contributions to the matrix element on the r.h.s. of
Eq.\anoma. The former are essentially given \gtw\
by diagrams where the mass term or the gluon condensate
couples to a pion (or generally a pseudoscalar meson) with the appropriate
quantum numbers, which then couples to the nucleon. A typical contribution to
the latter is that where the mass term couples directly to one of the nucleon's
quarks.

It is easy to see \gtw\ that the one-particle reducible contribution
to the matrix element of the mass term on the r.h.s. of Eq.\anoma\ displays
large isospin violation, because the singlet mass terms couples to the
triplet meson (the pion) with amplitude \gtw
\eqn\pion
{\langle 0| \left(\sum_i 2 m_i\,\bar\psi i\gamma_5\psi\right)|\pi^0\rangle =
{m_u-m_d\over m_u+m_d} f_\pi M_\pi^2,}
hence,
neglecting the $\eta$ contribution to the reducible portion of the matrix
element of Eq.\anoma, and
retaining only the pion one, leads immediately to isospin violation of order
${m_u-m_d\over m_u+m_d}$. Because
${m_d\over m_u}\approx 2$ \ref\wei{S.~Weinberg, in
``A Festschrift for I.~I.~Rabi, L.~Motz, ed. (New York Academy of Sciences, New
York, 1977)} this is a very large violation.
However, as brilliantly  shown in Ref.\gtw, the
Sutherland-Veltman theorem implies the vanishing of the one-pion matrix element
of the {\it full} r.h.s. of Eq.\anoma, which in turn implies that the one-meson
reducible matrix element of
the anomaly must display an isospin violation equal and opposite to that of the
mass term, such that the
overall reducible contribution to the matrix element of $\partial_\mu j^\mu_5$
is isospin conserving. This is borne out by explicit effective lagrangian
computations \venold. This  justifies the neglect of heavy quark
contributions to Eq.\corr, since the isospin violation of these is entirely
compensated by that of the anomaly \gtw.

What about the irreducible contributions? Despite the explicit isospin
violation from the (current) quark masses, the matrix element of the mass
term in a quark state with helicity $\lambda$ is perfectly isospin invariant:
\eqn\mass
{\langle q |m \bar\psi i\gamma_5\psi|q\rangle=2\lambda (p^\prime-p)\cdot s }
because its mass dependence cancels. Notice that this implies that the mass
term cannot be neglected even in the chiral limit; if this term is incorrectly
dropped from the Ward identities derived from Eq.\anoma\ one reaches
\ref\pirla{R.~L.~Jaffe and A.~Manohar, {\it Nucl. Phys.} {\bf B337}, 509
(1990)} the absurd
conclusion that the helicity of a
free quark vanishes in the chiral limit.
Eq.\mass\  leads thus to expect that no isospin violation should arise from
the irreducible matrix elements of the mass term.

It is interesting to consider instead irreducible contributions to the matrix
element of the anomaly from instanton-like field configurations. These have
been computed
in Ref.\ref\meshu{S.~Forte and E.~V.~Shuryak, {\it Nucl. Phys.} {\bf B357}, 153
(1991)} for a simplified model of a single massless quark, in the one-flavor
case. In that Ref., it was shown that the matrix element of the anomaly
receives a contribution from instanton-induced couplings which can be
identified with a contribution to $\Omega$ Eq.\twocomp, and which exactly
cancels the bare quark helicity, $\Delta q=- \Delta G$. This suggests that
the nonperturbative gluonic contribution $\Omega$
to the nucleon matrix element
of the axial current Eq.\twocomp\ may also be large and anticorrelated to
$\Delta q$, thus explaining the surprising experimental discovery \emc\
that this matrix element vanishes (within experimental uncertainties).

Such contribution arises from a direct coupling of the quark line to instantons
and is thus irreducible according to the above classification.  Now the
effective
instanton-induced chirality flipping interaction which generates it
arises due to fermionic zero modes in the instanton field, and is
thus suppressed by $1\over m_q$ if the quark is given a mass $m_q$.
This leads one to expect that in the many-flavor case the corresponding
contribution (which in the mean-field approximation is identical to the
one-flavor one \meshu) should display isospin violations of order
${\Omega^p\over\Omega^n}\sim {m_d\over m_u}$.

In order to get a feeling for the size of the expected violation of the sum
rule, we can very crudely assume  that $\Delta q_i$ take the
values obtained form the quark model and the Ellis-Jaffe sum rule (see \alta),
screened by $\Omega$ so that $A_1=0$, while
${\Omega^p\over\Omega^n}\sim {m_d\over m_u}$.
This leads to estimate the correction in Eq.\newbj\ to be
${5\over 18}(\Delta G^p-\Delta G^n)\sim 0.1$, which is roughly
one-half
of the Bjorken value ${1\over 6} {G_A\over G_V}$.
Otherwise stated, the extra
term on the r.h.s. of the Bjorken sum rule would thus be estimated to reduce
the Bjorken value by roughly a factor 2.

How would such a dramatic violation of the Bjorken sum rule affect the analysis
of data on the polarized proton structure function which has been presented
by the EMC collaboration \emc, and which has stirred so much attention in the
physics community? A minute's reflection reveals that it wouldn't affect it at
all. In order to extract from the data, which provide a value of the first
moment of the proton structure function $\int dx g_1(x)$, a value for the
singlet form factor $G_A$ Eq.\twocomp, which is of more direct
physical significance, one needs to use Eq.\isoe, expressed as
\eqn\isopr
{\Delta u - \Delta d={G_A\over G_V},}
together with an analogous relation for the SU(3) octet current. Now Eq.\isopr\
(and Eq.\isoe) are perfectly correct, regardless of whether the
assumption leading from Eq.\trip\ to Eq.\srule\ fails thereby
leading to violation of the
Bjorken sum rule. The existence of such a violation would only imply that,
due to the presence of the gluonic contribution in Eqs.\twocomp,\newbj,
knowledge of the proton  value of both the  first moment $A_1$  Eq.\ope
and the
singlet form factor $G_A$ Eq.twocomp is insufficient to determine the
respective neutron values.

Rather than providing evidence against QCD, a
violation of the Bjorken sum
rule would thus provide evidence in favor of the fact that indeed the matrix
element of the singlet axial
current receives an extra gluonic contribution, which is necessarily absent in
the matrix elements of the nonsinglet currents, since this is the only possible
source of isospin violation in the proton-neutron difference of structure
functions. Even though it will not provide a test of QCD, contrary to
widespread belief, the forthcoming experimental test of the Bjorken sum rule
will provide us with extremely valuable information on the nonperturbative
structure of this theory.

\bigskip
\noindent{\bf Acknowledgements:} I thank M.~Anselmino, M.~Arneodo,
and E.~Predazzi for discussions.

\vfill
\eject
\centerline{\bf REFERENCES}
\bigskip
\listrefs
\vfill
\eject
\bye